\documentclass[epj]{svjour}
\usepackage{graphics}
\begin{document}
\title{Nuclear dynamics in the EMC effect at Next to Next to Leading order}
\author{S.Atashbar Tehrani\inst{1,2}\thanks{\emph{Present address:} atashbar@ipm.ir}
\and H. Mouji\inst{3}
%
}                     
\offprints{}          
\institute{Department of Physics, Yazd Branch, Islamic Azad University, Yazd, Iran
\and School of Particles and Accelerators, Institute for
Research in Fundamental Sciences (IPM), P.O.Box 19395-5531,
Tehran, Iran \and Payame Noor University  Bushehr, Iran}
\date{Received: \today / Revised version: date}
%
\abstract{
We study in details the parameterizations of the nuclear parton distributions at the
next-to-next-to-leading order (NNLO) of $\alpha_s$. In low $x$ and $Q_0^2$,
we observe  negative gluon distribution at this order which
 signals  the saturation condition or  the quark-gluon plasma
condition. Our study also shows  the gluon distribution at (NNLO)  is less than next-to-leading order (NLO) of $\alpha_s$,
and the sea quark distribution at (NNLO) is larger than (NLO)
} 
\maketitle
\section{Introduction}
\label{intro}
Deep Inelastic Scattering (DIS) provides a tool for probing
the quark momentum distribution in the nucleons and in the
nuclei. Since the first indications that the DIS structure functions
measured in the charged-lepton scattering off the nuclei differ
significantly from those measured in the isolated nucleons,
there has been a continuous interest in fully understanding
the microscopic mechanism responsible in nuclei. This phenomena
called EMC effect which was discovered surprisingly in 1983 ~\cite{Aubert:1983xm}.After then it was being investigated  how it affects the momentum
distribution of quarks in nuclei. We indicate that the parton distributions in nuclei
are not simply as the parton densities in the nucleons. In
addition to the most commonly analyzed data sets for
deep-inelastic scattering of charged leptons off nuclei, we also
analysis  the Drell-Yan di-lepton production.
At low values of the Bjorken scaling variable $x$ the ratio is $R=F_2^A/F_2^D< 1$. 
At medium values of $x$, $R$ drops from $1$ to values as
low as $0.8$ and at large $x$ it reaches values larger than $1$.
While the latter feature can be quantitatively explained
by the smearing of the parton distribution functions arising
from the momentum distribution of nucleons in nuclei,
the former one is accounted for by including the
effect of nuclear shadowing.
This Phenomena divided to 4 region in $x \leq 0.05$ we have shadowing effect. The interval $ 0.05 \leq x\leq 0.3$ belongs to antishadowing
region. In $x = 0.6 \sim 0.7$ we encounter with  EMC effect and for large $x$ we have fermimotion.
While extended experimental and theoretical efforts
were put into studies of the origin of the EMC effect, an acceptable explanation has yet
to be found. One of the major goals of Quantum Chromo Dynamics (QCD) is the particular
investigation of the parton distribution of the proton and nuclei
which  for the first time has been observed by the European Muon
collaboration (EMC).
Deep inelastic scattering (DIS) experiment which have been
performed by NMC, SLAC, NMC, FNAL, BCDMS, HERMES and JLAB groups
\cite{Amaudruz:1995tq,Gomez:1993ri,Bodek:1983ec,Bodek:1983qn,Dasu:1988ru,Benvenuti:1987az,Ashman:1992kv,Ashman:1988bf,Adams:1992nf,Adams:1995is,Seely:2009gt,Arneodo:1995cs,Arneodo:1989sy,Bari:1985ga,Arneodo:1996rv,Arneodo:1996ru,Ackerstaff:1999ac} confirm  the specific feature of  nuclear
reaction at certain region of $x$-Bjorken variable which was
 first observed by EMC collaborations. This  specific feature  has
also been seen in Drell-Yan cross section ratios
\cite{Vasilev:1999fa,Alde:1990im}. In this paper we calculate the
nuclear parton distribution functions (NPDFs), using the global
analysis of experimental data, taking into account the ratio of
the structure function, $F_2^A/F_2^{A'}$, and Drell-Yan
cross-section ratios $\sigma_{DY}^{A}/\sigma_{DY}^{A'}$ by
employing the QCD-PEGASUS-package \cite{Vogt:2004ns}.

This paper consists of the following sections. In Sec.1, A
formalism to establish an analysis method to parameterize the
experimental data is introduced. In Sec.~\ref{Formalism} we embark this
analysis to follow our calculations. We give our calculations  in Sec.~\ref{The analysis of value}. The Hessian method is discussed in Sec.~\ref{hesian}
 and finally  results are given in Sec.~\ref{Results}.
\section{Formalism\label{Formalism}}
In order to calculate the parton distribution in nuclear media, we
need first the parton distributions in a free proton. We then use
a PDFs set which have been  parameterized at the input scale
$Q_0^2$=2 $GeV^2$ with the following standard form, quoted  from
\cite{JimenezDelgado:2008hf}:
\begin{equation}
xq(x,Q_0^2)=A_q x^{\alpha_q}(1-x)^{\beta_q}(1+\gamma_q x^{0.5}+\eta_q x)\nonumber\;.
\label{partonQ0}
\end{equation}
The PDFs in above are used as the valance quark distributions
$xu_v$, $xd_v$, the anti-quark distributions
$xs=\frac{x(\overline{u}+\overline{d}+\overline{s})}{3}$,
$x\Delta=x(\overline{d}-\overline{u})$, and  gluon distribution,
$xg$ .  In this
paper a comparison between the  results of   used  model in
\cite{JimenezDelgado:2008hf} and the experimental groups BCDMS, H1,
NMC, SLAC and ZEUS at an input scale of $Q_0^2=\mu^2_{NNLO}=2 GeV^2$ has been done.\\\\\\
\begin{figure}[tbh]
\resizebox{0.45\textwidth}{!}{\includegraphics{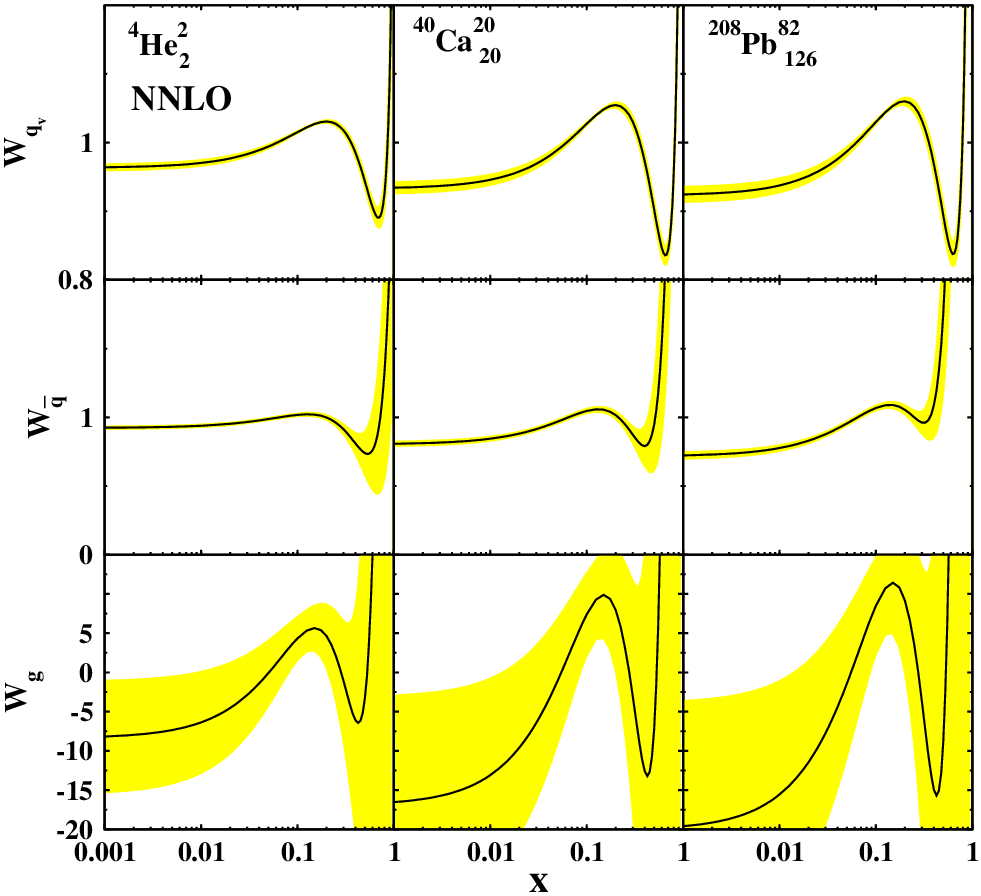}
}
\caption{The weight functions  for  $^4$He, $^{40}$Ca and
$^{208}Pb$ nuclei  at $Q_0^2$=2 $GeV^2$.}\label{fig:wHeCaPb1}
\end{figure}

We know the NPDFs are provided by a number of parameters at a
fixed $Q^2$  which are normally denoted by $Q_0^2$. The NPDFs are
related to PDFs in free proton and for this purpose nucleonic PDFs
are multiplied by a weight function $w_i$:
\begin{equation}
 f_i^A(x,Q_0^2)=w_i(x,A,Z)f_i(x,Q_0^2)\label{partonQ0A}\;.
\end{equation}
The parameters in weight function are obtained by a $\chi^2$
analysis procedure which are dependent on $x$ Bjorken variable, $A$ (mass number)
and $Z$ (atomic number).

Here we follow  the analysis given by
\cite{Hirai:2001np,Hirai:2004wq,Hirai:2007sx,Tehrani:2004hp,Tehrani:2006gy,Tehrani:2007hu,AtashbarTehrani:2012xh}
and assume the  functional form in below for  the weight function
in  Eq.~(\ref{partonQ0A}):\\\\\\\\
\begin{eqnarray}
 w_i(x,A,Z)&=&1+\left(1-\frac{1}{A^{\alpha}}\right)\nonumber\\
 &&\frac{a_i(A,Z)+b_i(A)x+c_i(A)x^2+d_i(A)x^3}{(1-x)^{\beta_i}}\;. \nonumber\\ \label{eq:waightfunction}
\end{eqnarray}
Combining the weight function in Eq.~(\ref{eq:waightfunction})
with PDFs  in Eq.~(\ref{partonQ0}), will yield us  us NPDFs as in
what follows:
\begin{eqnarray}
u_v^A(x,Q_0^2)&=& w_{u_v}(x,A,Z)\frac{Z\; u_v(x,Q_0^2)+ N\; d_v(x,Q_0^2)}{A}\nonumber \;,\\
d_v^A(x,Q_0^2)&=& w_{d_v}(x,A,Z)\frac{Z\; d_v(x,Q_0^2)+ N\; u_v(x,Q_0^2)}{A}\nonumber\;, \\
\overline{u}^A(x,Q_0^2)&=& w_{\overline{q}}(x,A,Z)\frac{Z\; \overline{u}(x,Q_0^2)+ N\; \overline{d}(x,Q_0^2)}{A}\nonumber \;,\\
\overline{d}^A(x,Q_0^2)&=& w_{\overline{q}}(x,A,Z)\frac{Z\; \overline{d}(x,Q_0^2)+ N\; \overline{u}(x,Q_0^2)}{A}\nonumber \;,\\
s^A(x,Q_0^2)&=& w_{\overline{q}}(x,A,Z)s(x,Q_0^2)\nonumber \;,\\
g^A(x,Q_0^2)&=& w_{g}(x,A,Z)g(x,Q_0^2) \;.\label{partonQW}
\end{eqnarray}
In the first four equations, $Z$ term as atomic number  indicates
the number of protons  and the $(N=A-Z$) term indicate the
number of neutrons in in nuclei while  the $SU(3)$ symmetry  is
apparently broken there. If the number of protons and neutrons in
a nuclei are equal to each other (iso-scalar nuclei) such as
$^2D$,$^4He$,$^{12}C$ and $^{40}Ca$ nuclei, the  valence quarks
$u_v^A$ and $d_v^A$ and $\overline{u}^A$ and $\overline{d}^A$
would have similar distributions. In the case that $Z$ and $A$
numbers are not equal in the nuclei, it can be concluded that
anti-quark distributions
$(\overline{u}^A,\overline{d}^A,\overline{s}^A)$ in the nuclei
would not be equal to each other
\cite{AtashbarTehrani:2012xh,Kumano:1997cy,Garvey:2001yq}. For the strange quark
 distributions in the nuclei  some research
studies are still  being done \cite{Kusina:2012vh} but we assume
the common case in which it is assumed $(s=\overline{s})$. In
Eq.~(\ref{eq:waightfunction}) we take $\alpha=1/3$ as in
\cite{Sick:1992pw} considering nuclear volume and surface
contributions. In addition there are three constraints for the
parameters which are existed in  Eq.~(\ref{eq:waightfunction})
namely the nuclear charge $Z$, baryon number  (mass number)$A$ and
momentum conservation
\cite{Hirai:2001np,Hirai:2004wq,AtashbarTehrani:2012xh,Frankfurt:1990xz}  as in
following:
\begin{eqnarray}
 Z&=&\int  \frac{A}{3}\left[2u_v^A-d_v^A\right](x,Q_0^2)\;dx\;,\nonumber\\
 3&=&\int \left[u_v^A+d_v^A\right](x,Q_0^2) \;dx\;,\nonumber\\
 1&=&\int x\left[u_v^A+d_v^A+2\left\{\overline{u}^A+\overline{d}^A+\overline{s}^A\right\}\right.\nonumber\\
 &&\left. g^A\right](x,Q_0^2) \;dx \;.\label{baryon&moment}
\end{eqnarray}

\begin{figure*}[htb]
\vspace{0.53cm}
\centering
\hspace*{0.0cm}
\begin{minipage}{.56\textwidth}
\resizebox{0.56\textwidth}{!}{\includegraphics{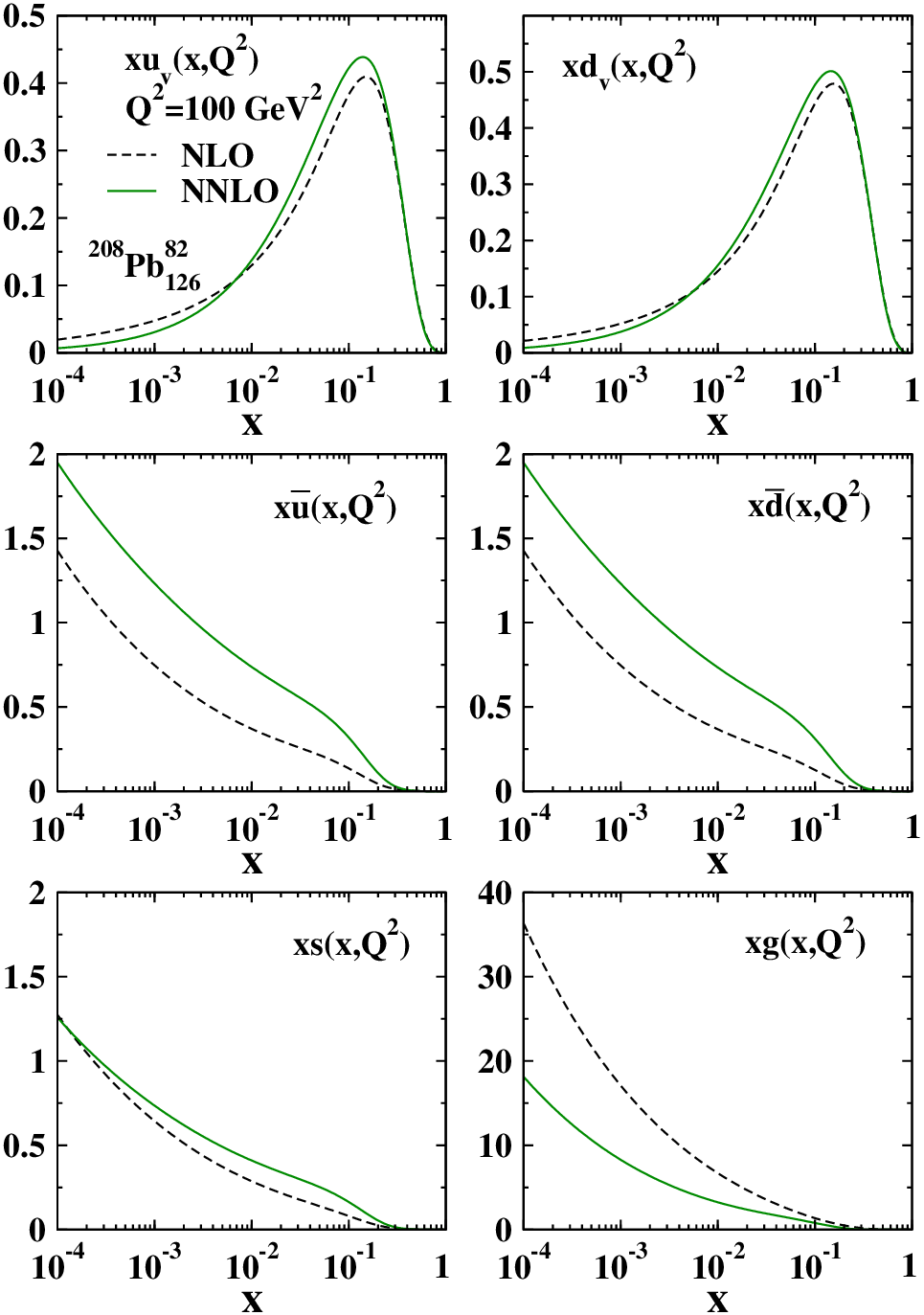}}

  \hspace*{0.0cm}
\end{minipage}%
\begin{minipage}{.56\textwidth}
  \resizebox{0.56\textwidth}{!}{\includegraphics{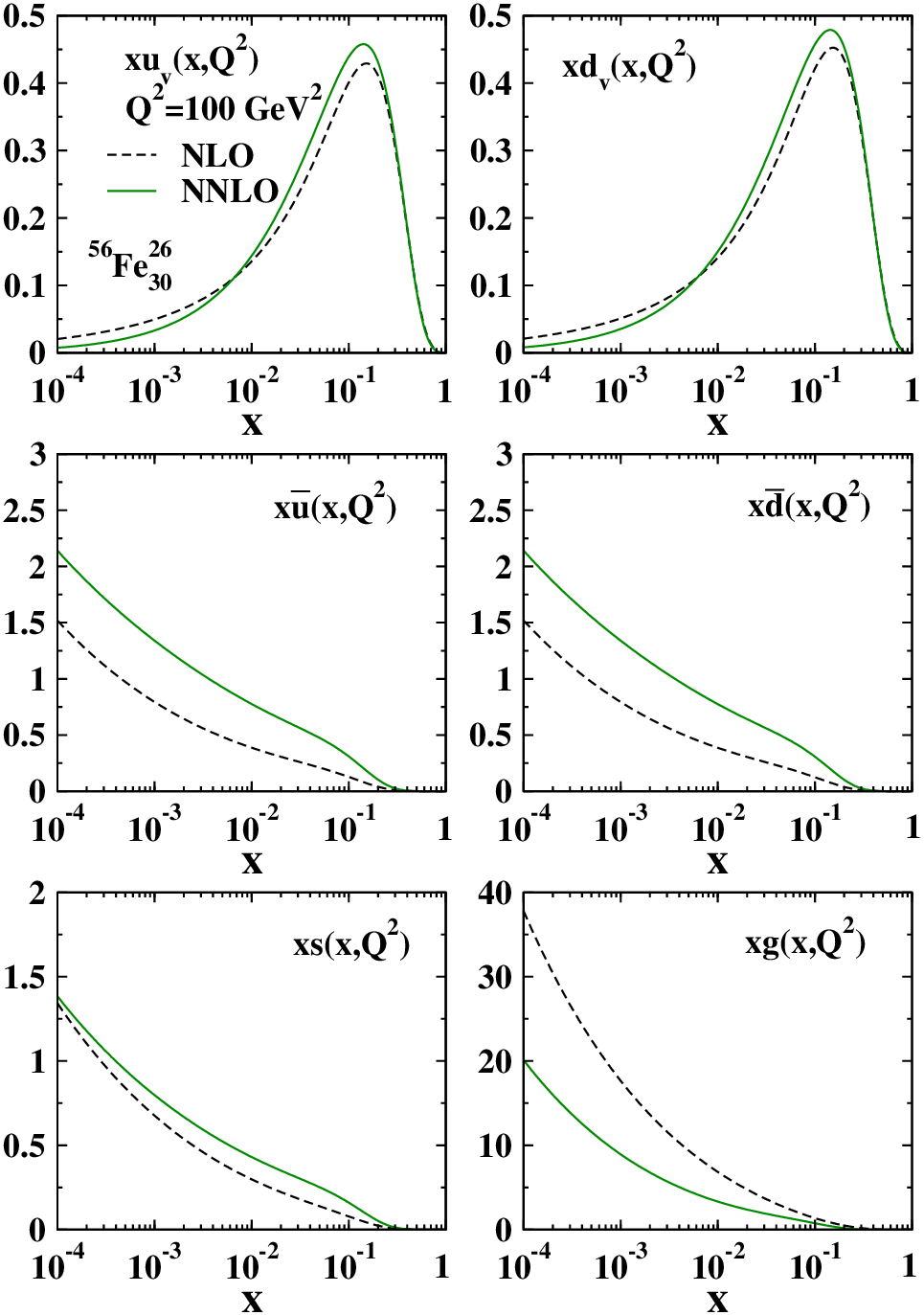}}

\end{minipage}
\caption{Parton distribution for Lead and Iron at $Q^2$=100 $GeV^2$
in NNLO and its comparison with the NLO reslts. \cite{AtashbarTehrani:2012xh}.} \label{fig:PbQ100}
\end{figure*}

\begin{figure*}[htb]
\vspace{0.65cm}
\centering
\hspace*{0.0cm}
\begin{minipage}{.56\textwidth}
\resizebox{0.56\textwidth}{!}{\includegraphics{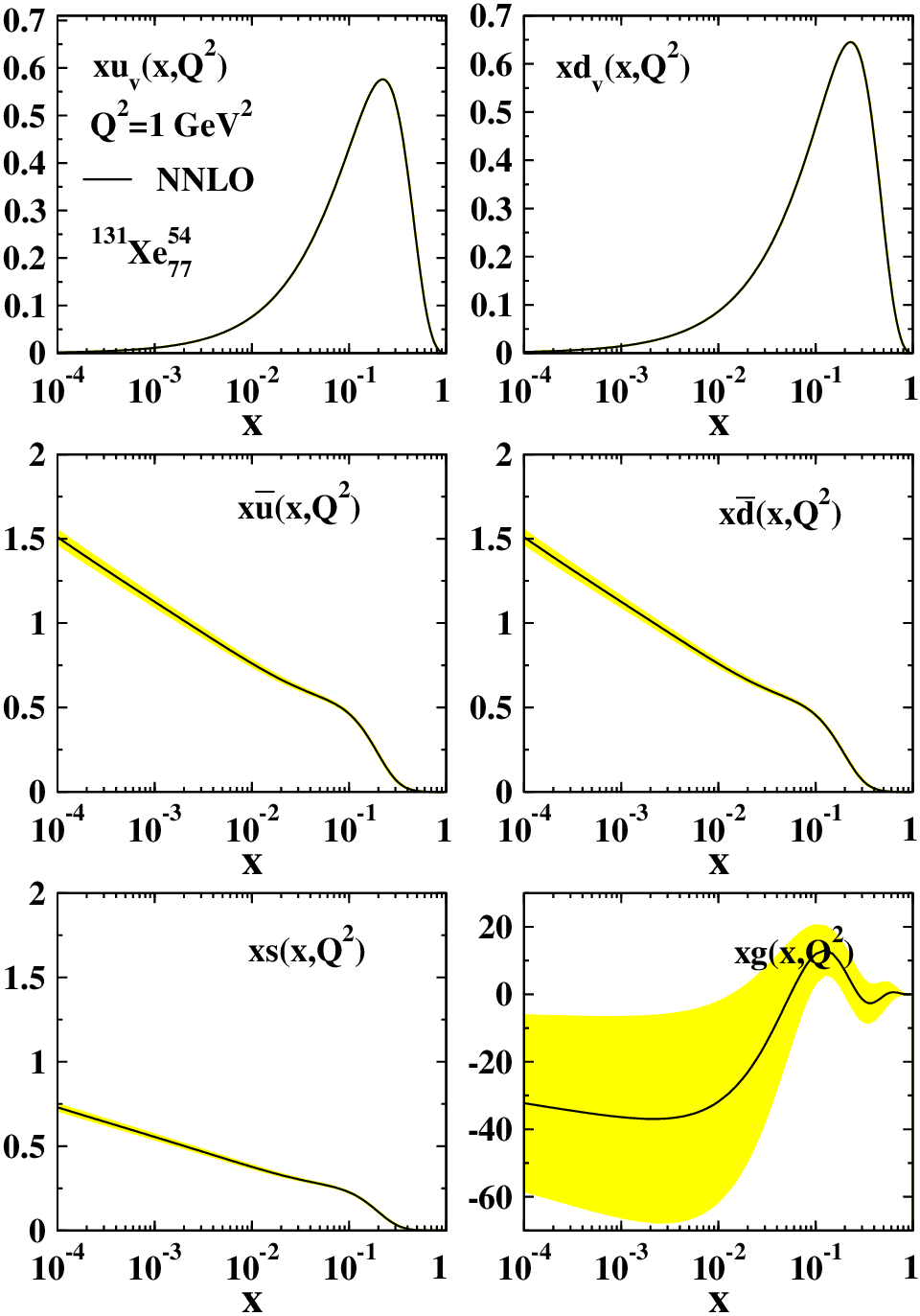}}
  \hspace*{0.0cm}
\end{minipage}%
\begin{minipage}{.56\textwidth}
 \resizebox{0.56\textwidth}{!}{\includegraphics{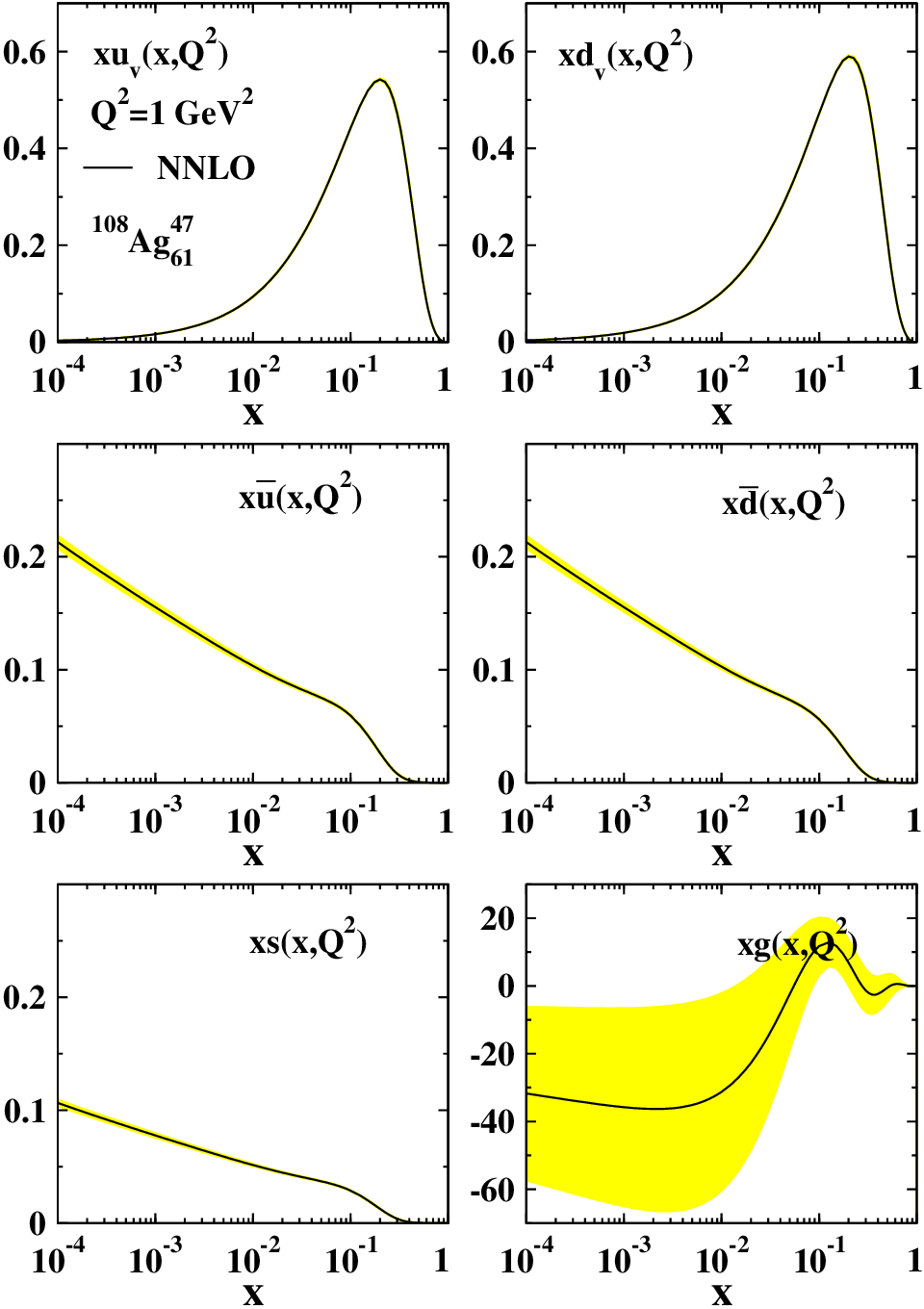}}
\end{minipage}
\caption{Parton distribution for Xenon and Silver at $Q^2=$ 1 $GeV^2$
in NNLO approximation including the  error band.} \label{fig:XeQ1}
\end{figure*}

 \section{Overview of the available experimental data\label{The analysis of value}}
In Table ~\ref{table1}, we listed a number of the data which has
been prepared by different experimental groups for the
$F_2^A/F_2^D$ ratio. In this ratio the numerator is denoting the
structure function of a nuclei and denominator is representing
the structure function of Deuterium. The total number of data for
the ratio in which the numerator includes nuclei like Helium
($He$), Lithium ($Li$) and etc. is  equal to 1079. In Table
 the number of $F_2^A/F_2^{A^{\prime}}$ ratio for
$Be/C$, $Al/C$, $Ca/C$, $Fe/C$, $Sn/C$, $Pb/C$, $C/Li$ is 308. For
Drell-Yan cross section ratios in table ~\ref{table1} the number
of data is equal to 92 while the related ratio are $C/D$, $Ca/D$,
$Fe/D$, $W/D$, $Fe/Be$ and $W/Be$. In the employed analysis the
total number of data is 1479. The interval range of $Q^2$ values
is $Q^2 \geq 1 GeV^2$ and the smallest value for Bjorken variable
$x$ is equal to 0.0055.

\begin{table}
\begin{center}
\begin{tabular}{cccc}
\hline\hline Nucleus & Experiment & \#of data & Reference \\
\hline\hline
(F$_{2}^{A}$/F$_{2}^{D}$) &  &  &  \\
He/D & SLAC-E139 & 18 &  \cite{Arneodo:1989sy}\\
& NMC-95 & 17 &  \cite{Amaudruz:1995tq}\\
Li/D & NMC-95 & 17 & \cite{Amaudruz:1995tq} \\
Li/D(Q$^{2}$dep.) & NMC-95 & 179 & \cite{Amaudruz:1995tq} \\
Be/D & SLAC-E139 & 17 & \cite{Gomez:1993ri} \\
C/D & EMC-88 & 9 &  \cite{Ashman:1988bf}\\
& EMC-90 & 5 & \cite{Arneodo:1989sy} \\
& SLAC-E139 & 7 &  \cite{Gomez:1993ri}\\
& NMC-95 & 17 & \cite{Amaudruz:1995tq} \\
& FNAL-E665 & 5 & \cite{Adams:1995is} \\
& JLAB-E03-103 & 103 & \cite{Seely:2009gt} \\
C/D(Q$^{2}$dep.) & NMC-95 & 191 & \cite{Amaudruz:1995tq}  \\
N/D & BCDMS-85 & 9 & \cite{Bari:1985ga} \\
& HERMES-03 & 153 &  \cite{Ackerstaff:1999ac}\\
Al/D & SLAC-E49 & 18 & \cite{Bodek:1983ec} \\
& SLAC-E139 & 17 & \cite{Gomez:1993ri} \\
Ca/D & EMC-90 & 5 & \cite{Arneodo:1989sy} \\
& NMC-95 & 16 &   \cite{Amaudruz:1995tq}\\
& SLAC-E139 & 7 & \cite{Arneodo:1989sy} \\
& FNAL-E665 & 5 & \cite{Adams:1995is} \\
Fe/D & SLAC-E87 & 14 & \cite{Bodek:1983qn} \\
& SLAC-E139 & 23 &\cite{Gomez:1993ri}  \\
& SLAC-E140 & 10 & \cite{Dasu:1988ru} \\
& BCDMS-87 & 10 & \cite{Benvenuti:1987az} \\
Cu/D & EMC-93 & 19 &\cite{Ashman:1992kv}  \\
Kr/D & HERMES-03 & 144 & \cite{Ackerstaff:1999ac}  \\
Ag/D & SLAC-E139 & 7 &  \cite{Gomez:1993ri}\\
Sn/D & EMC-88 & 8 & \cite{Ashman:1988bf} \\
Xe/D & FNAL-E665-92 & 5 & \cite{Adams:1992nf} \\
Au/D & SLAC-E139 & 18 & \cite{Gomez:1993ri} \\
& SLAC-E140 & 1 & \cite{Dasu:1988ru} \\
Pb/D & FNAL-E665-95 & 5 & \cite{Adams:1995is} \\ \hline\hline
(F$_{2}^{A}$/F$_{2}^{A^{\prime }}$) &  &  &  \\
Be/C & NMC-96 & 15 & \cite{Arneodo:1996ru} \\
Al/C & NMC-96 & 15 & \cite{Arneodo:1996ru} \\
Ca/C & NMC-96 & 24 & \cite{Amaudruz:1995tq} \\
& NMC-96 & 15 & \cite{Arneodo:1996ru} \\
Fe/C & NMC-96 & 15 &  \cite{Arneodo:1996ru}\\
Sn/C & NMC-96 & 146 &  \cite{Arneodo:1996ru}\\
     & NMC-96 & 15  &  \cite{Arneodo:1996ru}\\
Pb/C & NMC-96 & 15 &  \cite{Arneodo:1996ru}\\
C/Li & NMC-95 & 24 &   \cite{Amaudruz:1995tq} \\
Ca/Li & NMC-95 & 24 &  \cite{Amaudruz:1995tq}  \\ \hline\hline
($\sigma _{DY}^{A}/\sigma _{DY}^{A^{\prime }})$ &  &  &  \\
C/D & FNAL-E772-90 & 9 & \cite{Alde:1990im} \\
Ca/D & FNAL-E772-90 & 9 & \cite{Alde:1990im} \\
Fe/D & FNAL-E772-90 & 9 & \cite{Alde:1990im} \\
W/D & FNAL-E772-90 & 9 & \cite{Alde:1990im} \\
Fe/Be & FNAL-E866/NuSea & 28 & \cite{Vasilev:1999fa} \\
W/Be & FNAL-E866/NuSea & 28 & \cite{Vasilev:1999fa} \\ \hline
Total &  & 1479 &  \\ \hline
\end{tabular}
\caption[]{Different experimental results for the
F$_{2}^{A}$/F$_{2}^{D}$ , F$_{2}^{A}$/F$_{2}^{A^{\prime}}$ and $\sigma _{DY}^{A}/\sigma _{DY}^{A^{\prime }}$
ratio at $Q^2\geq1.0$ GeV$^2$. Number of
data points and the related references are also listed.}
\label{table1}
\end{center}
\end{table}

\begin{figure*}
\begin{center}
\resizebox{0.78\textwidth}{!}{\includegraphics{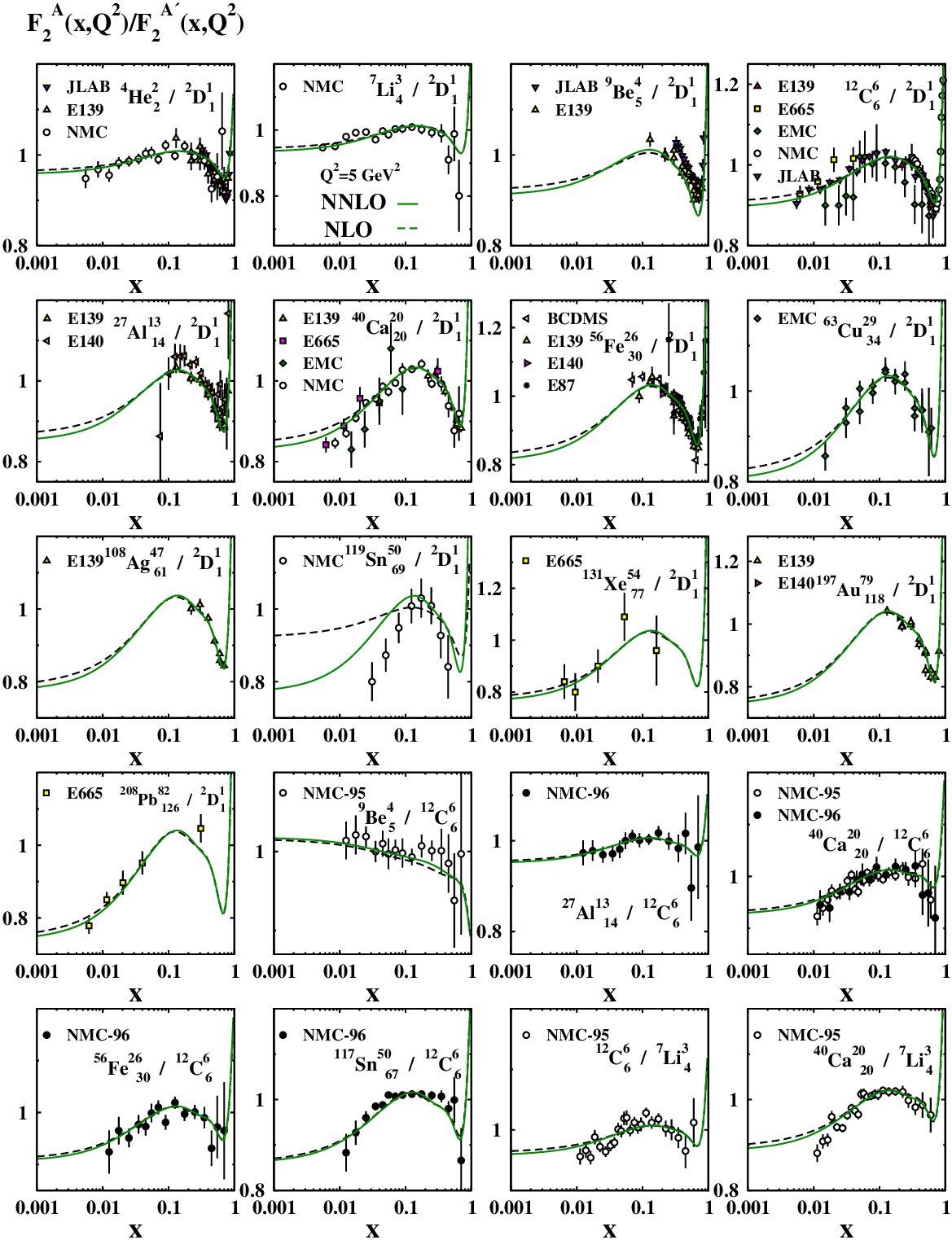}}
\caption[]{EMC ratio for several nuceli at $Q^2=$ 5 $GeV^2$,
in NNLO approximation and its comparison with the  NLO approximation \cite{AtashbarTehrani:2012xh} .}\label{fig:emc}
\end{center}
\end{figure*}

\section{The analysis of $\chi^2$ value}
\label{hesian}
We use the MINUIT fitting package \cite{MINUIT} to fit the
experimental data, including the structure function
F$_{2}^{A}$/F$_{2}^{A^{\prime }}$ and Drell-Yan cross section
ratios.

The optimized value of total $\chi^2$ is defined by
\begin{equation}
\chi^2=\sum_j\frac{(R_j^{data}-R_j^{theo})^2}{(\sigma_j^{data})^2}\;.\label{chi}
\end{equation}
This relation yield us the  proper parameters for NPDFs. Here
$R_j^{data}$  indicates the experimental values for
F$_{2}^{A}$/F$_{2}^{A^{\prime }}$ or $\sigma _{DY}^{A}/\sigma
_{DY}^{A^{\prime }}$ ratio  and $R_j^{theo}$ is denoting the
theoretical result for  the parameterized NPDFs. In our
calculations, we take $Q_0^2$=2 $GeV^2$ \cite{JimenezDelgado:2008hf}
and the  $\chi^2$ analysis  is done, based on the DGLAP evolution
equations \cite{Vogt:2004ns}. Our calculations are done in the
next-next-to-leading (NNLO) approximation in which the modified minimal
subtraction scheme $(\overline{MS})$  is used \cite{Gluck:1989ze}.

Therefore the nuclei structure function is written as in
following:
\begin{eqnarray}
 F_2^{A}(x,Q^2)&=&\sum_{i=u,d,s}e_i^2x\left[1+
 a_sC^1_{q}(x)+a_s^2C^2_{q}(x)\right]\otimes(q_i^A+\overline{q}_i^A) \nonumber\\
 &&+\frac{1}{2f}(a_sC^1_g(x)+a_s^2C^2_g(x))\otimes xg\;.\label{F2A}
\end{eqnarray}

In this equation, $C^{1,2}_{q,g}$ are wilson coefficient in NLO and NNLO approximation
\cite{vanNeerven:1999ca,vanNeerven:2000uj} and the symbol $\otimes$ denotes the convolution
integral:
\begin{equation}
f(x)\otimes
g(x)=\int_x^1\frac{dy}{y}f\left(\frac{x}{y}\right)g(y)\;.\label{covoloution}
\end{equation}
We employ the CERN program library MINUIT to minimize the $\chi^2$
value. Following that  an error analysis can be done, using the
Hessian matrix. The NPDFs uncertainties are estimated, using the
Hessian matrix as in following:
\begin{equation}
        [\delta f^A(x)]^2=\Delta \chi^2 \sum_{i,j}
          \left( \frac{\partial f^A(x,\xi)}{\partial \xi_i}
                \right) _{\xi=\hat{\xi}}
          H_{ij}^{-1}
          \left( \frac{\partial f^A(x,\xi)}{\partial \xi_j}
                \right) _{\xi=\hat{\xi}}  ,
        \label{hessian}
\end{equation}
where $H_{ij}$ is the Hessian matrix, $\xi_i$ is a quantity
refereing to  the parameters which exist in NPDFs and $\hat\xi$
indicates the amount of the parameter which makes  an extremum
value for the related derivative\cite{AtashbarTehrani:2012xh}.\\\\\\

\begin{figure}[htb]
\begin{center}
\resizebox{0.3\textwidth}{!}{\includegraphics{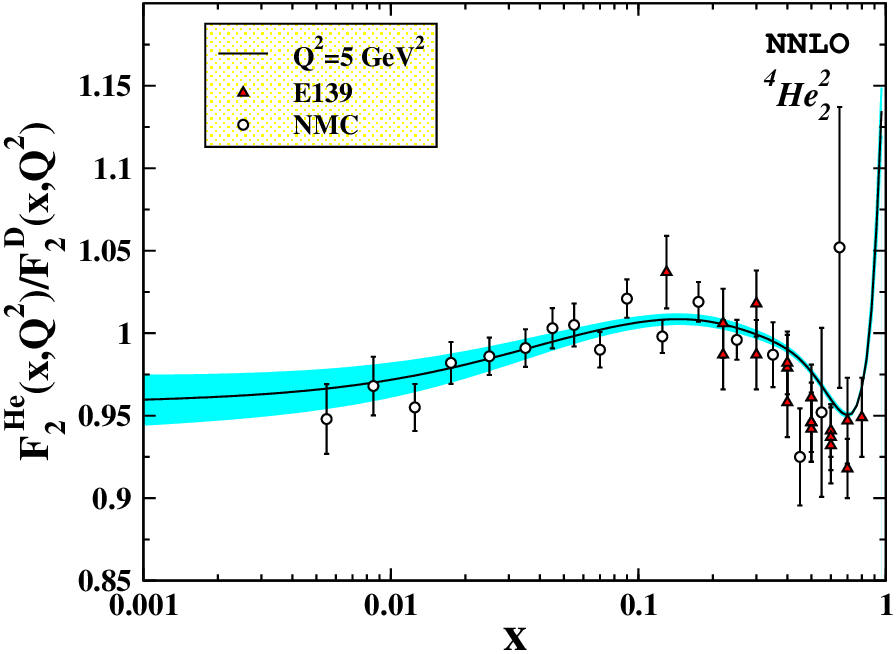}}
\caption{EMC effect for Helium at $Q^2=$ 5 $GeV^2$,
in NNLO approximation including the  error band.}\label{fig:HeQ5}
\end{center}
\end{figure}

\begin{figure}[htb]
\begin{center}
\resizebox{0.3\textwidth}{!}{\includegraphics{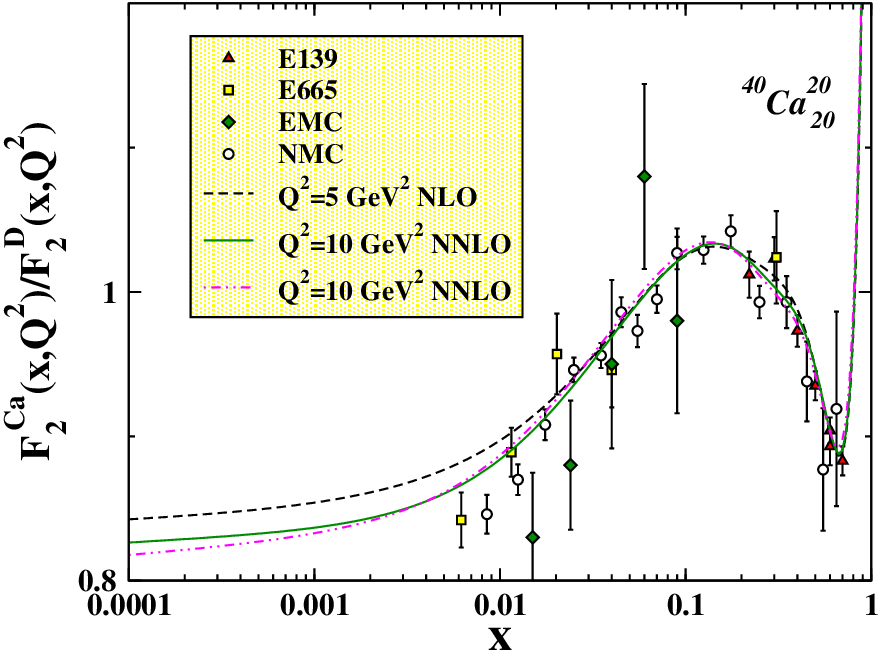}}
\caption{EMC effect for Calcium at $Q^2=$ 5,10 $GeV^2$,
in NNLO approximation and its comparison with the NLO approximation \cite{AtashbarTehrani:2012xh} .}\label{fig:CaQ510}
\end{center}
\end{figure}

\begin{figure}[htb]
\begin{center}
\resizebox{0.35\textwidth}{!}{\includegraphics{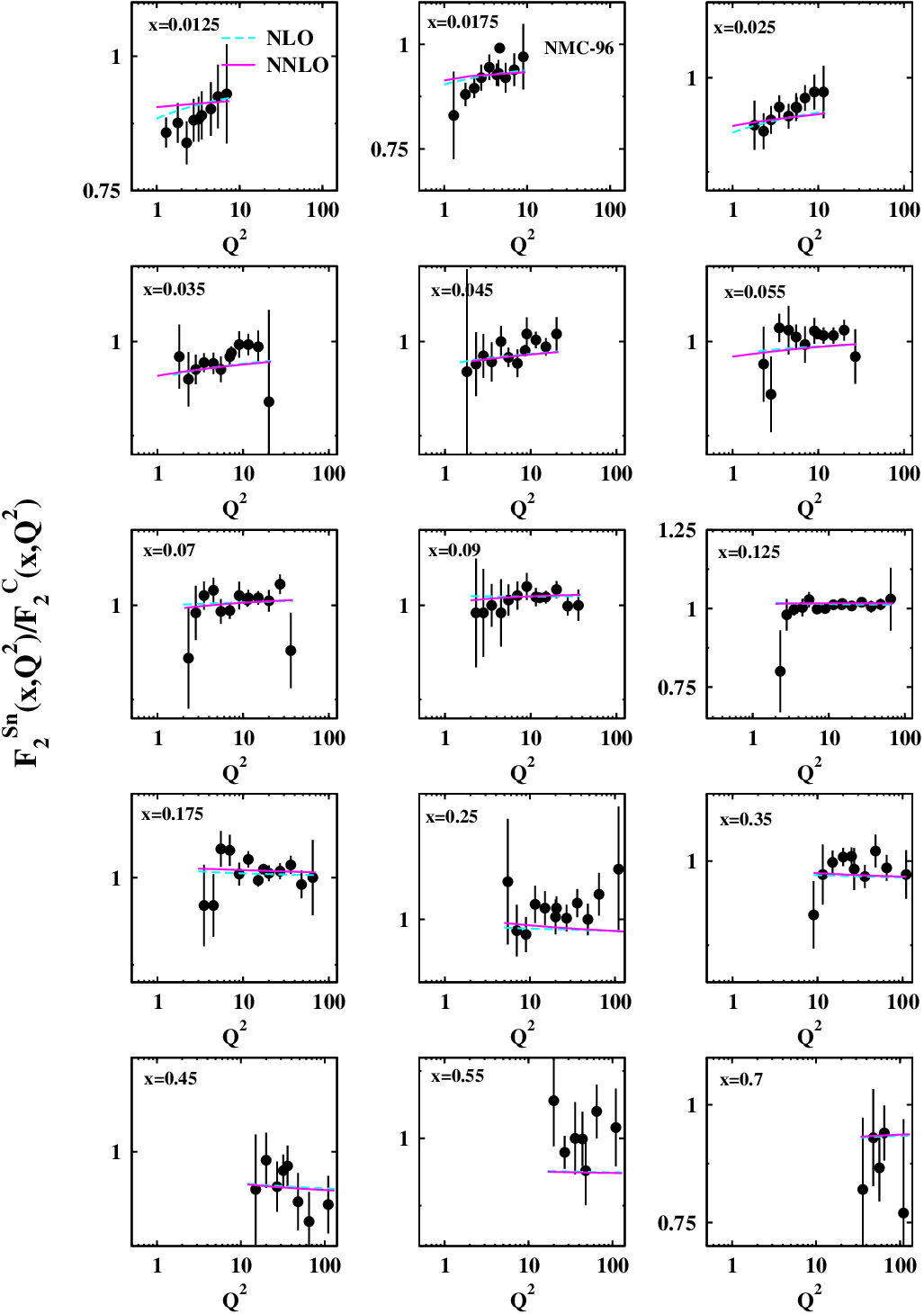}}
\caption{$Q^2$ dependence of $F_2^{Sn}/F_2^C$\cite{Arneodo:1996ru} in comparison  with
the NNLO  and NLO approximations \cite{AtashbarTehrani:2012xh}.}\label{fig:EMCSntoC}
\end{center}
\end{figure}

\section{Results}
\label{Results}
In our analysis we use the weight function method as in
Ref. \cite{AtashbarTehrani:2012xh}. However in this analysis we assume that
the coefficients in Eq.~(\ref{eq:waightfunction}) depend on
nuclear mass number, $A$.  In our analysis which
is done in the NNLO approximation, we get to  $\chi^2/D.O.F=
1614.90/1462=1.1046$ . The number of the data points
for the nuclei and Drell-Yan ratios is totaly $1479$. The  number
of parameters which is used in our fitting procedure is equal to
$17$.  In Fig.~\ref{fig:wHeCaPb1} we depict the weight functions for the
Helium, Calcium and Lead nuclei at initial value $Q_0^2$= 2
$GeV^2$. In this figure, the uncertainties are indicated by the
bands. We extract the NPDFs for Lead and Iron nuclei at $Q^2$= 100
GeV$^2$ in NNLO model and compare it with NLO result in \cite{AtashbarTehrani:2012xh}
Fig.~\ref{fig:PbQ100}. These figure show that the sea quark
distribution in NNLO approximation is greater than NLO and also the gluon distribution in
NNLO approximation is less than NLO approximation.

\begin{table*}
\small
\begin{center}
\begin{tabular}{lll}
\hline\hline
$a_{v}$ & $a_{\overline{q}}$ & $a_{g}$ \\
Appendix A & ($-0.176\pm 5.78\times 10^{-3})A^{0.123\times 10^{-2}\pm }$ &
Appendix \\
$b_{v}$ & $b_{\overline{q}}$ & $b_{g}$ \\
$(2.052\pm 2.58\times 10^{-2})A^{-1.89\times 10^{-2}\pm 2.91\times 10^{-3}}$
& $(3.82\pm 8.09\times 10^{-2})A^{0.128\pm 3.91\times 10^{-3}}$ & $570.3\pm
58.5$ \\
$c_{v}$ & $c_{\overline{q}}$ & $c_{g}$ \\
$(-6.769\pm 3.68\times 10^{-2})A^{-1.17\times 10^{-2}\pm 1.29\times 10^{-3}}$
& $(-18.11\pm 0.41)A^{0.145\pm 4.63\times 10^{-3}}$ & $-2575.5\pm 271.2$ \\
$d_{v}$ & $d_{\overline{q}}$ & $d_{g}$ \\
$(5.165\pm 3.63\times 10^{-2})A^{3.613\times 10^{-3}\pm 1.78\times 10^{-3}}$
& $(16.12\pm 0.887)A^{0.239\pm 9.54\times 10^{-3}}$ & $2985.03\pm 437.7$ \\
$\beta _{v}$ & $\beta _{\overline{q}}$ & $\beta _{g}$ \\
$0.4$ Fixed & $0.1$ Fixed & $0.1$Fixed\\ \hline\hline
\end{tabular}
\caption[]{Parameters obtained by  analyzing the weight functions
for valance quark, sea quark and gluon distributions
.}\label{table4}
\end{center}
\end{table*}

\begin{figure*}[htb]
\vspace{0.95cm}
\centering
\hspace*{0.0cm}
\begin{minipage}{.69\textwidth}
\resizebox{0.69\textwidth}{!}{\includegraphics{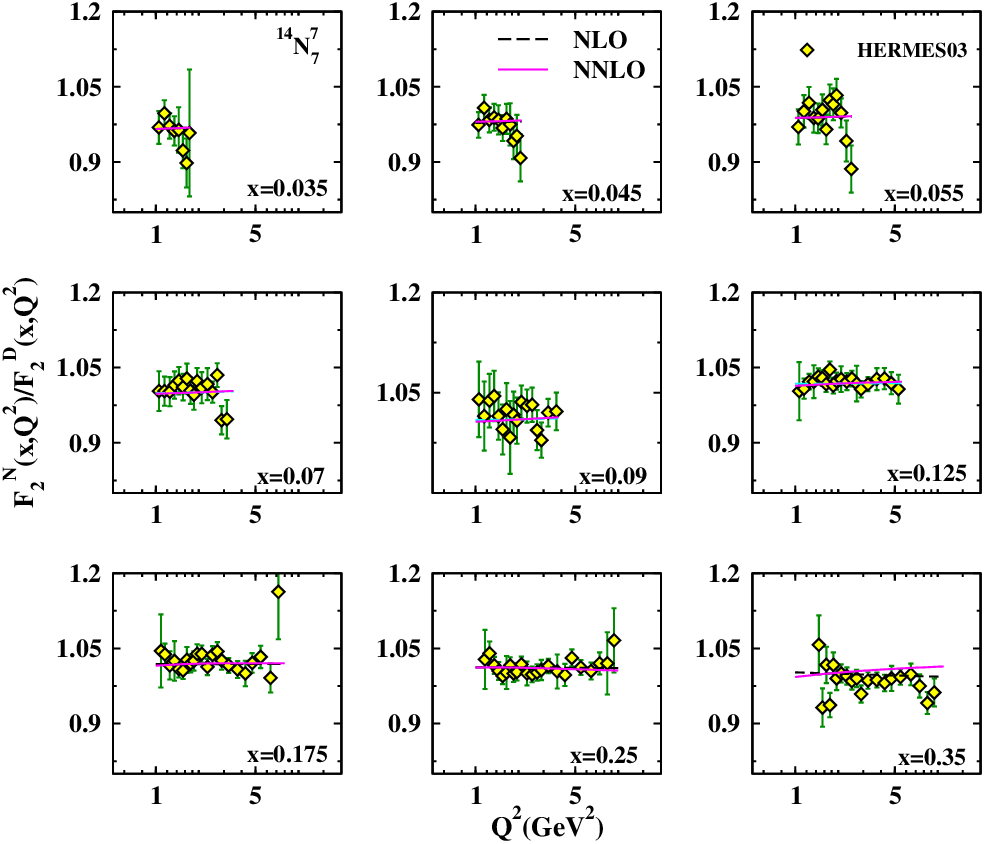}}
\end{minipage}%
\hspace*{-2.50cm}
\begin{minipage}{.69\textwidth}
 \resizebox{0.69\textwidth}{!}{\includegraphics{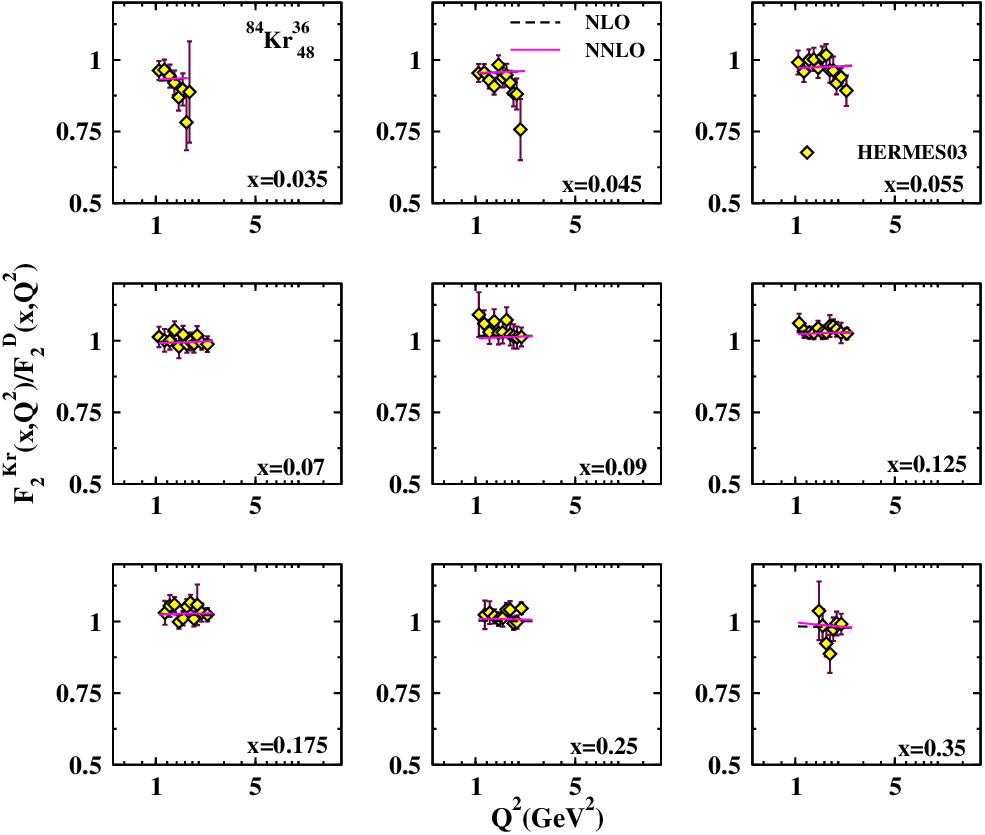}}
\end{minipage}
\caption{$Q^2$ dependence of $F_2^{Kr}/F_2^D$ and $F_2^{N}/F_2^D$ \cite{Arneodo:1996ru} in  comparison with the results of 
NLO  and NNLO approximations \cite{AtashbarTehrani:2012xh}.}\label{fig:emckrd}
\end{figure*}

In Fig.~\ref{fig:XeQ1} we plot the parton distributions in $Q_0^2=2 GeV^2$
with error bar for Xenon and Silver. In this Figure, we have negative distribution  for gluon.
We compare in Fig.~\ref{fig:emc} our theoretical results with
the related DIS data.
In Fig.~\ref{fig:HeQ5} we show the EMC effect with uncertainties and  compare with
experimental data in $Q^2=5 GeV^2$. In Fig.~\ref{fig:CaQ510} we compare the theoretical
model in $Q^2=5 , 10$ in NNLO,   plot NLO approximation ~\cite{AtashbarTehrani:2012xh} and
compare them with experimental data. We get good result and better than NLO approximation.
In Figs ~\ref{fig:EMCSntoC} and ~\ref{fig:emckrd}, we depict the ratios of $F_2^{Sn}/F_2^D$,
$F_2^{Kr}/F_2^D$  and $F_2^{N}/F_2^C$ with respect to $Q^2$
values and compare them with the available experimental data
\cite{Arneodo:1996ru,Ackerstaff:1999ac} in NLO and NNLO approximation. The nuclear parton
distribution functions and their uncertainties are determined by
analyzing  the $F_2$ and Drell-Yan experimental data. The
uncertainties are again estimated by the Hessian method.

\begin{figure}[htb]
\begin{center}
\vspace{0.95cm}
\resizebox{0.4\textwidth}{!}{\includegraphics{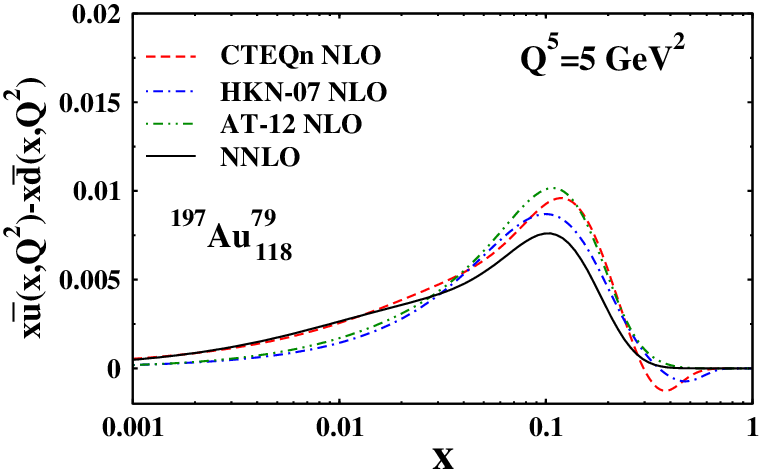}}
\caption{Flavor symmetry $x{\overline u}-x{\overline d}$ in Gold
a at $Q^2$=5 GeV$^2$ and its comparison  with
HKN-07\cite{Hirai:2007sx},n-CTEQ\cite{Schienbein:2009kk} and \cite{AtashbarTehrani:2012xh}
results.}\label{fig:ubar-dbar}
\end{center}
\end{figure}

Flavor symmetry in nuclei such as $^2$D,$^4$He,$^{12}$C and
$^{40}$Ca are like each other in which ${\overline u}={\overline
d}= s$. For another nuclei that the number of their protons and
neutrons are not equal,  we have  the  $SU(3)$ flavor symmetry
breaking. We compare our model with assumption ${\overline
u}\ne{\overline d}\ne s$ with respect to HKN-07
\cite{Hirai:2007sx} ,n-CTEQ \cite{Schienbein:2009kk} and AT-12 \cite{AtashbarTehrani:2012xh}
 results. This analysis has been done for Gold in Fig.~\ref{fig:ubar-dbar}
at $Q^2$=5 GeV$^2$ in which we have $SU(3)$  symmetry breaking.
 In the first step $20$ parameters have been optimized by minimizing the $\chi^2$ value
and in the second one since we fixed three parameters $\beta
_{v}$, $\beta _{\overline{q}^{A}}$ and $\beta _{g}$, we just need
to determine $17$ parameters of the  weight functions via our
fitting procedure. The reason that we have to fix these three
parameters is that to control the fermi motions of the partons
inside the nuclei at the large values of $x$. For the weight
functions of the valance and sea quark distributions, we choose an
A-dependent function while the weight function for the gluon
distribution is assumed independent of A number. The numerical
values in Table. ~\ref{table4}, are listed on this base. The
parameters $a_{u_v}$, $a_{d_v}$ and $a_g$  are fixed by the three
sum rules, given by Eq.~(\ref{baryon&moment}).


If we intend to discuss about  this analysis
at some small enough value of $x$ the number of gluons distributions is negative and we have saturation condition in
$Q_0^2$ for small $x$ values. Does this mean there are no gluons in that region?
No, it means we are in saturation region ~\cite{d'Enterria:2006nb} in nuclei where  
overlapping of gluons with smaller impact parameter is increased. It is  due to the increasing number of nucleons along a
straight line. Consequently   the probability for gluon recombination effects inside the  nucleus would be increased ~\cite{Chiu:2012ak}. 
This  leads to gluon saturation and  negative behavior at low x region.  We should also notice that
the small-$x$ behavior of singlet NNLO splitting functions is negative for small values of $Q^2$ at low $x$
region ~\cite{Vogt:2004mw,Vogt:2004gi}.  The negative distribution for gluon in low $Q^2$ shows that we
have saturation condition that is very important for nuclei-nuclei collisions and for studying the quark-gluon plasma condition that
occurs in the Big-bang or in the neutrons stars.
In hight-energy $AA$ collisions, hard or semi-hard parton scattering in the initial stage may result in a large amount of
jet production. In particular, the multiple minijets whose typical transverse moment is a few GeV could give rise to an
important fraction of the transverse energy produced in the heavy ion collisions \cite{Kajantie:1987pd,Eskola:1988yh} in the RHIC and LHC .

\newpage
\section*{Appendix A}
Having three sum rules which give us the nuclear charge $Z$,
baryon number $A$ and momentum conservation as in
Eq.(\ref{baryon&moment}), we can calculate  the three parameters
$a_{u_v}(A,Z)$, $a_{d_v}(A,Z)$ and $a_g(A,Z)$:
\begin{eqnarray}
a_{u_v}(A,Z)&=&-\frac{Z I_1(A)+(A-z) I_2(A)}{Z I_3+(A-Z) I_4},\nonumber\\
a_{d_v}(A,Z)&=&-\frac{Z I_2(A)+(A-z) I_1(A)}{Z I_4+(A-Z) I_3},\nonumber\\
a_g(A,Z)&=&-\frac{1}{I_8}\left\{a_{u_v}(A,Z)\left[\frac{Z}{A} I_5+\left(1-\frac{Z}{A}\right) I_6\right]
\right.\nonumber\\
&+& \left. a_{d_v}(A,Z)\left[\frac{Z}{A} I_6+\left(1-\frac{Z}{A} I_5\right)\right]+I_7(A)\right\}\;.\nonumber\\
\label{au,ad,ag}
\end{eqnarray}
To obtain the numerical values for these parameters in any nuclei,
we need to  calculate the following integrals \cite{Hirai:2007sx}:
\begin{eqnarray}
I_1(A)&=& \int  \frac{H_v(x,A)}{(1-x)^{\beta_v}}u_v(x)\;dx\;,
\nonumber \\
I_2(A)&= &\int  \frac{H_v(x,A)}{(1-x)^{\beta_v}}d_v(x)\;dx\;,
\nonumber \\
I_3&=& \int  \frac{1}{(1-x)^{\beta_v}}u_v(x)\;dx\;,
\nonumber \\
I_4&= &\int  \frac{1}{(1-x)^{\beta_v}}d_v(x)\;dx\;,
\nonumber \\
I_5&=& \int  \frac{x}{(1-x)^{\beta_v}}u_v(x)\;dx\;,
\nonumber \\
I_6&=& \int  \frac{x}{(1-x)^{\beta_v}}d_v(x)\;dx\;,
\nonumber \\
I_7(A)&=&\int x \biggl[ \frac{H_v (x,A)} {(1-x)^{\beta_v}}
                              \{u_v(x)+ d_v(x)\}
\nonumber \\
&&+ \frac{a_{\overline{q}}(A)+H_{\overline{q}}(x,A)}{(1-x)^{\beta_{\overline{q}}}}
   2 \{ \overline{u}(x)+\overline{d}(x)+\overline{s}(x) \}\nonumber\\
&&+\frac{H_g(x,A)}{(1-x)^{\beta_g}}g(x) \biggr]\;dx\;,
\nonumber \\
I_8&=&\int  \frac{x}{(1-x)^{\beta_g} }g(x)\;dx\;,\label{I18}
\end{eqnarray}
where $\beta_v=0.4$,  $\beta_{\bar q}=\beta_g=0.1$ and $H_i(x,A)$
is given by
\begin{eqnarray}
H_v(x,A)&=&b_v(A) x+c_v(A) x^2+d_v(A) x^3,\nonumber\\
H_{\overline{q}}(x,A)&=&b_{\overline{q}}(A) x+c_{\overline{q}}(A) x^2+d_{\overline{q}}(A) x^3,\nonumber\\
H_g(x)&=&b_g+c_gx^2+d_gx^3 \;.\label{Hvqg}
\end{eqnarray}

The results of the eight integrals in above, depend on  the atomic
number and  are different for each nuclei.

\section*{Appendix B }\label{App:B}
The \texttt{FORTRAN} package containing our unpolarized structure
functions, $F_{2}^A(x,Q^{2})$, for nuclei, as well as the
unpolarized parton densities $xu_{v}^A(x,Q^{2})$,
$xd_{v}^A(x,Q^{2})$, $xs^A(x,Q^{2})$, $x\bar{u}^A(x,Q^{2})$,
$x\bar{d}^A(x,Q^{2})$, $xg^A(x,Q^{2})$ and their uncertainties at
NNLO approximation in the $\overline{{\rm MS}}$--scheme  can be
obtained via e-mail from the author. In this package we assumed
$10^{-4}\leq x\leq 0.999$ and $2\leq Q^2\leq 10^{5}$ GeV$^2$.



%
%

\end{document}